\documentclass[aps,prl,twocolumn,showpacs]{revtex4-1}
\usepackage{amssymb}
\usepackage{graphicx}
\usepackage{amsmath}

\usepackage{times}
\usepackage{color}
\usepackage{subfigure}
\usepackage{setspace}
\usepackage{bm}

\begin{document}

\newcommand {\beq} {\begin{equation}}
\newcommand {\eeq} {\end{equation}}
\newcommand {\bqa} {\begin{eqnarray}}
\newcommand {\eqa} {\end{eqnarray}}
\newcommand {\ba} {\ensuremath{b^\dagger}}
\newcommand {\Ma} {\ensuremath{M^\dagger}}
\newcommand {\psia} {\ensuremath{\psi^\dagger}}
\newcommand {\psita} {\ensuremath{\tilde{\psi}^\dagger}}
\newcommand{\lp} {\ensuremath{{\lambda '}}}
\newcommand{\A} {\ensuremath{{\bf A}}}
\newcommand{\Q} {\ensuremath{{\bf Q}}}
\newcommand{\kk} {\ensuremath{{\bf k}}}
\newcommand{\qq} {\ensuremath{{\bf q}}}
\newcommand{\kp} {\ensuremath{{\bf k'}}}
\newcommand{\rr} {\ensuremath{{\bf r}}}
\newcommand{\rp} {\ensuremath{{\bf r'}}}
\newcommand {\ep} {\ensuremath{\epsilon}}
\newcommand{\nbr} {\ensuremath{\langle ij \rangle}}
\newcommand {\no} {\nonumber}
\newcommand{\up} {\ensuremath{\uparrow}}
\newcommand{\dn} {\ensuremath{\downarrow}}
\newcommand{\rcol} {\textcolor{red}}

\begin{abstract}
In the ionic Hubbard model, the onsite repulsion $U$, which drives a
Mott insulator and the ionic potential $V$, which drives a band
insulator, compete with each other to open up a window of charge
fluctuations when $U \sim V$. We study this model on square and cubic lattices in the
limit of large $U$ and $V$, with $V\sim U$. Using an effective Hamiltonian and a
slave boson approach with both doublons and holes, we find that the system undergoes a phase transition as a function of $V$ from an
antiferromagnetic Mott insulator to a paramagnetic insulator with
strong singlet correlations, which is driven by a condensate of
``neutral'' doublon-hole pairs. On further increasing $V$, the
system undergoes another phase transition to a superconducting phase driven
by condensate of ``charged'' doublons and holes. The superfluid phase,
characterized by presence of coherent (but gapped) fermionic
quasiparticle, and $hc/e$ flux quantization, has a high $T_c \sim t $ which shows a dome shaped behaviour as a function of
$V$. The paramagnetic insulator phase has a deconfined U(1) gauge
field and associated gapless photon excitations. We also discuss how these phases can be detected in the ultracold
atom context.
 \end{abstract}
\title{ Superconductivity from Doublon Condensation in the Ionic Hubbard Model}
\author{Abhisek Samanta and Rajdeep Sensarma}
 \affiliation{Department of Theoretical Physics, Tata Institute of Fundamental
 Research, Mumbai 400005, India.}

\date{\today}

\maketitle

A dramatic observable effect of strong interactions between fermions on a
lattice  is the formation of Mott insulating states, where charge
motion is suppressed due to large on-site
repulsion~\cite{Mott,Imada_rev}. This effect 
occurs in a large class of materials like transition metal
oxides~\cite{Mott_ins_expt1,Mott_ins_expt2,Mott_ins_expt3,Mott_ins_expt4}, including 
parent compounds of cuprate high $T_c$ superconductors~\cite{Mott_ins_expt3,Mott_ins_expt4}. Recently, Mott insulators have been observed in systems of ultracold
fermions on optical lattices~\cite{MI_cold1,MI_cold2}, where the repulsive Fermi Hubbard model
with tuneable Hamiltonian parameters can be implemented faithfully.

A theoretically challenging problem is to ascertain the fate of a system in
proximity to a Mott
insulator, where charge fluctuations are induced by different means;
e.g. by doping the system away from commensurate filling (high $T_c$
cuprate superconductors)~\cite{Lee_rev,Plain_vanilla} or by changing ambient pressure
(organic superconductors)~\cite{Org_Sc} or simply by changing
the ratio of the interaction energy scale to the kinetic energy scale
(ultracold atomic systems)~\cite{Cold_atom_cf}.  Experimentally, when charge fluctuation
is induced around a Mott insulator, competing order parameters lead
to a very rich phase diagram~\cite{cuprate_comp_order,Org_Sc} with an ubiquitous presence of
superconducting phases~\cite{randARPES,cuprate_Sc, Org_Sc}. 

The Ionic Hubbard model is defined on bipartite lattices~\cite{Hub_Torr}, where, in addition to the kinetic energy
($\sim t$)
and the local Hubbard repulsion ($\sim U$), the fermions are affected
by a constant one-body potential difference between the two
sublattices ($\sim V$). This model, originally proposed
to explain ionic to neutral transitions ~\cite{Hub_Torr,Nagaosa}, has also been used to
describe ferroelectric transitions ~\cite{ihm_ferro1,ihm_ferro2,ihm_ferro3}. It
has recently been implemented in the context of ultracold
atoms~\cite{Esslinger_ihm} where the relative strengths of $U$ and $V$ can be tuned controllably. In the absence of
interactions, this model describes a band insulator at
half-filling due to doubling of the unit cell, while, in
the limit of strong interactions and weak potential, the system
goes into the Mott insulating phase. While both $V$ and $U$, by themselves, promote insulating behaviour,
they compete with each other leading to a window of charge
fluctuations when they are comparable to each other.

In this Letter, we study the ionic Hubbard model in the limit of large
$U$ and $V$, with $U\sim V$. 
The ionic Hubbard model has been studied in the literature using
various techniques like exact diagonalization ~\cite{ihm_ed}, DMFT
~\cite{ihm_DMFT1,ihm_DMFT2,ihm_DMFT3,ihm_DMFT4,ihm_DMFT5} and DMRG~\cite{ihm_ed,Manmana}. Most of these works have
focused on the regime $U\sim t$, where they have found an ionic to
neutral transition in 1D and a metallic phase between a band insulator
and a Mott insulator. In contrast, we will focus on $U \gg
t$, so that we approach a charge fluctuation regime starting from a
Mott insulator. For large $U/t$, Manmana {\it et. al}~\cite{Manmana}
have studied the model in 1D using DMRG, while a slave boson approach has been
used in the limit of small $V/U$~\cite{ihm_sbmft}.

We use a new canonical transformation to derive an effective
dimer-dipole Hamiltonian for $U/t$, $V/t$ $\gg 1$, with $U\sim V$, and study its phase diagram at
$T=0$ at half-filling within a slave
boson mean field theory.
Our key results are: 
(i) Fermions hop by converting a spin-singlet on a bond to a charge
dipole, with a doublon and a hole on the two sublattices, inducing charge fluctuation. (ii) Since the kinetic energy
prefers spin-singlets, the antiferromagnetic order decreases with
increasing $V$ and vanishes at a critical potential $V_{c1}$. (iii)
Beyond a critical $V_{c2}$ the doublons and holes forming the dipole
on the bond delocalize, leading to their Bose condensation. This
creates a superconducting state with $hc/e$ vortices. (iv) The
superfluid stiffness and critical temperature of this phase
shows a non-monotonic dome shaped behaviour as a function of $V$. (v) At large
$U$, $V_{c2}>V_{c1}$ and the intervening phase is a paramagnetic
insulator described by strong singlet fluctuations and a paired
superfluid~\cite{paired_SF1,paired_SF2,paired_SF3} of doublon-hole pairs. This phase shows
deconfinement of gauge degrees of freedom, which enforce projection
constraints in the system, and associated emergent gapless
``photons''~\cite{deconfined}. At lower values of $U$, the system shows a first order
transition from an AF insulator to a superconductor.

{\bf Low Energy Effective Hamiltonian:} The ionic Hubbard Hamiltonian
is given by $H=H_0+H_T$,
\bqa
\no H_0&=&U\sum_i n_{i\up}n_{i\dn} +\frac{V}{2}\sum_i (-1)^i n_i\\
H_T&=&-t\sum_{\langle ij\rangle\sigma}c^\dagger_{i\sigma}c_{j\sigma}=
\sum_nT^n_{-}+T^n_{+}
\label{ihham}
\eqa
Here $H_0$, the local part of the Hamiltonian, includes
the ionic potential $\pm V/2$ on $B(A)$ sublattice. $T^n_{+(-)}$ hops a Fermion
 from the $A(B)$ to the $B(A)$ sublattice and
increases the double occupancy by $n$, with $n=0,\pm 1$, causing an energy change $\Delta
E^n_s=nU+Vs$. In the regime, $U\gg t$, $V\gg t$, $U-V\sim zt$, where
$z$ is the co-ordination number, $T^1_-$ and $T^{-1}_{+}$ are low
energy hoppings, even at half-filling. They
create/annihilate a doublon-hole pair, i.e. a charge dipole  on a bond, so that the Hubbard repulsion is offset by the
potential energy gained in the process. 
The canonical transformation~\cite{Macdonald,smref1} then eliminates all high energy hopping processes
and we obtain the low-energy
effective Hamiltonian,
\beq
 \tilde{H}= H_0 +T^1_-+T^{-1}_++\frac{1}{U+V} [T_+^1,T_-^{-1}]+ \frac{1}{V} [T_+^0,T_-^0]
\label{eq:effham}
\eeq
We note that this effective Hamiltonian is obtained by an expansion
around the $U=V$ limit, and is notably different from the effective
Hamiltonian obtained by perturbing around the $U\gg t, V\sim 0$
limit~\cite{otherham}. Our effective Hamiltonian does
not contain terms $\sim 1/(U-V)$, i.e. the resonant processes encountered in
the expansion around $V=0$ are treated non-perturbatively (as ${\cal
  O}(t)$ hopping terms) in our
approach. We will later see that this new hopping process kills
antiferromagnetism and leads to a superconductivity of doublons and
holes in this regime.  The second order terms lead to spin-spin and
density-density interactions, as well as intra-sublattice hopping
terms. We will now use a slave boson mean-field
theory to determine the phase diagram of this effective model.

{\bf Slave Boson Formalism:}
In the slave boson formalism~\cite{slave_boson}, 
the fermion operator $c^\dagger_{i\sigma}=f^\dagger_{i\sigma}h_i+\sigma
f_{i\bar{\sigma}}d^\dagger_i$, where $f_{i\sigma}$ is a
spin $1/2$ chargeless fermion (spinon), and  holes $h_i$ and doublons
$d_i$ are spinless bosons carrying opposite charge,
$\pm 1$. The physical Hilbert space is obtained by imposing the constraint $
d^\dagger_id_i+h^\dagger_ih_i+\sum_\sigma
f^\dagger_{i\sigma}f_{i\sigma}=1$ at every lattice site.
\begin{figure}[t]
\includegraphics[width= \linewidth]{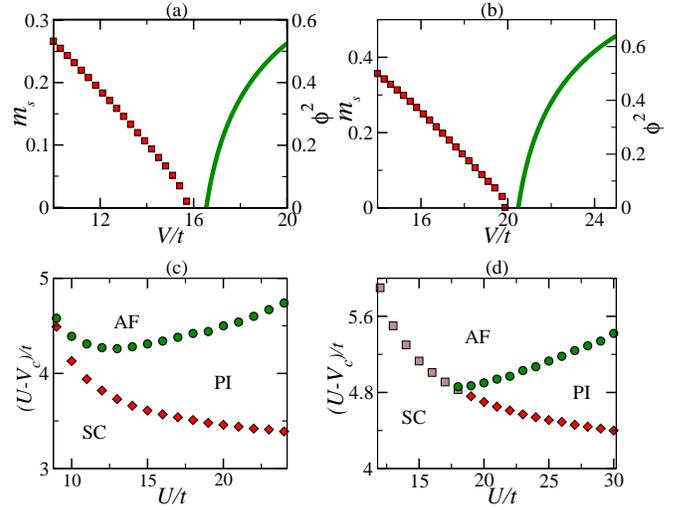}
\caption{ The staggered magnetization $m_s$ and the condensate fraction
  of the doublons (holes) $\phi^2$ as a function of the ionic
  potential $V$ for (a) square lattice with $U=20t$ and (b) cubic
  lattice with $U=25 t$. (c) and (d):
  the phase diagram in the $U-V$ plane for (c) square lattice and
  (d) cubic lattice.}
\label{fig:mzphi}
\end{figure}

At low energies, the doublons are
projected out of the $B$ sublattice and holes are projected out of
$A$ sublattice as they cost an energy  $ \sim U$~\cite{footnote1}. 
The low
energy degrees of freedom are $f_{iA(B)\sigma }$, $d_{iA}$
and $h_{iB}$, with the
constraints $d^\dagger_{iA}d_{iA}+\sum_\sigma
f^\dagger_{iA\sigma }f_{iA\sigma }=1$  and $h^\dagger_{iB}h_{iB}+\sum_\sigma
f^\dagger_{iB\sigma }f_{iB\sigma }=1$ to be implemented by Lagrange
multipliers $\mu^A$ and $\mu^B$ respectively.
The effective Hamiltonian is
\bqa
\no \tilde{H}&=& \sum_i \mu^d_i n^d_{iA}+\mu^h_in^h_{iB}+\mu^f_{iA}n^f_{iA}+\mu^f_{iB}n^f_{iB}\\
 & -t&\sum_{\langle ij\rangle \sigma} \sigma
f_{jB\overline{\sigma}}f_{iA\sigma}d^\dagger_{iA}h^\dagger_{jB}+h.c\\
\no & +&\frac{2t^2}{U+V}\sum_{\langle ij\rangle}
\left[\vec{S}_i\cdot\vec{S}_j-\frac{1}{4} n^f_in^f_j\right]+\frac{2t^2}{V}\sum_{\langle ij\rangle} n^d_{iA}n^h_{jB} 
\label{eq:slbos}
\eqa
where $\mu^d_i=U-V-2\mu-\mu^A_i$, $\mu^h_i=-\mu^B_i$,
$\mu^f_{iA}=-V/2-\mu-\mu^A_i$ and $\mu^f_{iB}=V/2-\mu-\mu^B_i$, and
$\mu$ is the chemical potential.
The low energy ${\cal O}(t)$ hopping term is a process which converts a spinon-singlet on a
bond to a charge dipole (doublon-hole pair) on the bond and vice
versa. The super-exchange interaction has a reduced scale of
$2t^2/(U+V) \sim t^2/U$, while there is a nearest neighbour repulsion ${\cal O}(t^2/V)$ between
a doublon and a hole~\cite{smref1}. 

\begin{figure}[t]
\includegraphics[width=\linewidth]{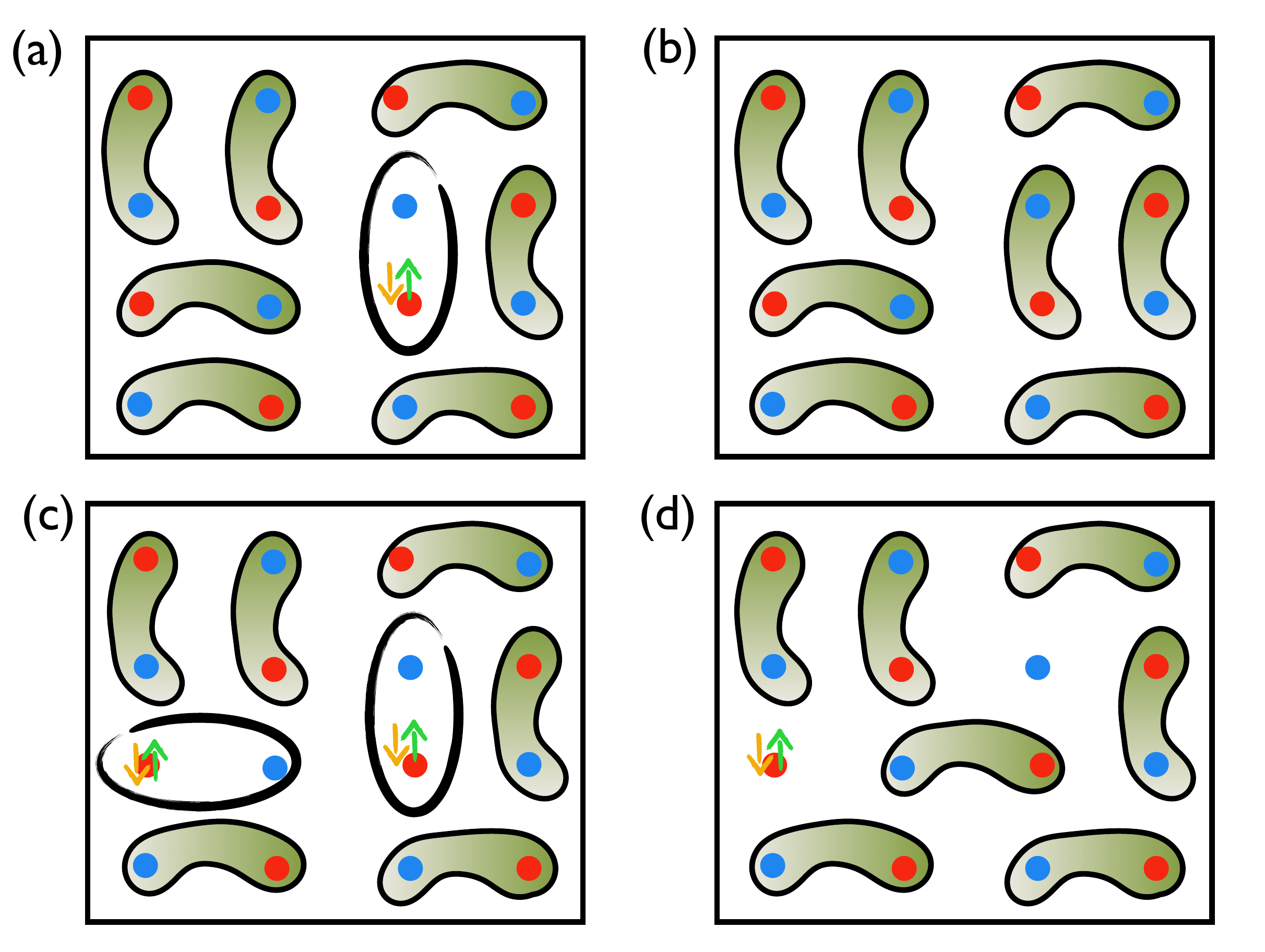}
\caption{Schematic showing delocalization of doublons and
  holes. Red (blue) dots denote $A(B)$ sublattice, and the green bonds are spin-singlets: (a) A dipole is created in the background of singlets (b)
  At low density of dipoles, it fluctuates back to a singlet (c) At
  high density of dipoles a neighbouring bond can fluctuate to create
  a $d-h$ pair (d) The doublon of one dipole and the hole of another
  dipole creates a singlet, leaving a delocalized doublon
  and hole.}
\label{fig:deloc}
\end{figure}

{\bf Mean Field Theory and the Phase Diagram:} 
We first treat the effective Hamiltonian within a mean field theory, where the constraints are
maintained on the average. We give mean field expectation value to
staggered magnetization $m_s= \sigma\tau \langle f^\dagger_{i\tau\sigma}f_{i\tau\sigma}\rangle$,
where $\tau=\pm 1$ for $A(B)$ sublattice, the doublon hole pairing
amplitude $c_1=\langle
d_{iA}h_{jB}\rangle$ and the spinon singlet amplitude $c_2=\sigma \langle
f_{iA\sigma}f_{jB\overline{\sigma}}\rangle$, while the condensation of 
individual doublons/holes are indicated by the condensate fraction $\phi^2$. 
\begin{figure}
\includegraphics[width=\linewidth]{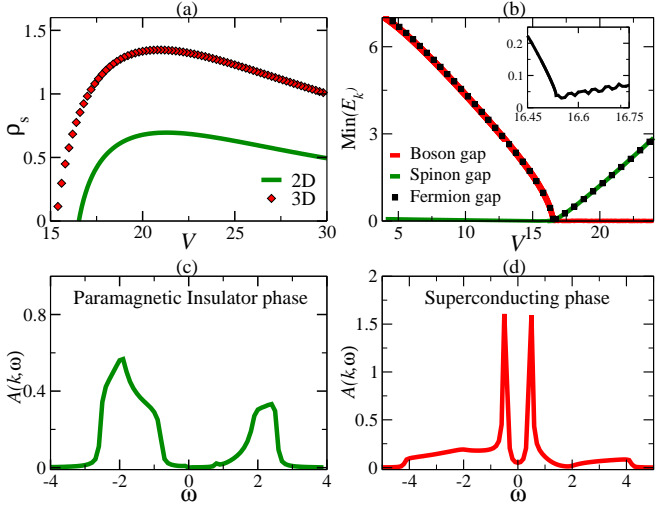}
\caption{(a) The superfluid stiffness as a function of $V$ for the
  square and the cubic lattice. (b) The single particle fermion gap together with the
  spinon and the doublon/hole gap as a function of $V$. The inset
  shows that the gap always remains finite. (c) and (d): ${\cal A}(k,\omega)$ as a function of energy
  $\omega$ for $k=[\pi/2,\pi/2]$ in (c) the paramagnetic insulator
  and (d)
the superconducting phase.}
\label{fig:spectral}
\end{figure}
There are three distinct phases that are obtained within the
slave-boson mean-field theory: (i) an antiferromagnetic (AF) Mott insulator
at small $V$ (ii) a superconducting (SC) phase with a Bose condensate of linear
combination of doublons and holes when $V\sim
U$ and (iii) an intervening paramagnetic insulator (PI) which is a paired 
superfluid of doublon-hole pairs. 
A charge order exists in
all the phases, but it does not correspond to any spontaneously broken
symmetry for the ionic Hubbard model. 

The AF order
parameter $m_s$ is shown as a function of $V$ in Fig.~\ref{fig:mzphi}
for large for (a) square  ($U=20 t$) and a
(b) cubic lattice ($U=25t$). It decreases monotonically
with $V$ and vanishes at $V=V_{c1}$ through a weakly first order
transition. This can be understood from the fact that the kinetic
energy favours spin-singlets which fluctuate to form charge dipoles on
the bond and the energy cost of forming these dipoles is decreasing
with increasing $V$. The subsequent dynamics of doublons
and holes are explained schematically in Fig.~\ref{fig:deloc}. If the
density of doublon-hole pairs are low, the
dipoles fluctuate back to spin singlets
before they can delocalize, leading to a paired superfluid with local
charge fluctuations, as shown in Fig.~\ref{fig:deloc} (a) and (b). If
the density of dipoles is high, and there are
dipoles on two neighbouring bonds, the
doublon of one dipole and the hole of the other dipole can fluctuate
back to a singlet leaving a separated doublon and hole, as shown in
Fig.~\ref{fig:deloc} (c) and (d). We note that this process can be easily visualized in the cold
atom context through a quantum gas microscope~\cite{Cold_atom_cf}, which
measures the local number parity in the system. In the PI phase, the
even parity sites should occur in pairs, while the condensed phase
would have a large number of isolated even parity sites.
Fig.~\ref{fig:mzphi} (a) and (b) also shows the condensate
fraction $\phi^2$ as a function of $V$, which is finite beyond a $V_{c2} > V_{c1}$,  leading to
a SC phase of delocalized condensed charged bosons. In the region
$V_{c1} < V < V_{c2}$, the system is a paramagnetic insulator with short
range spinon singlets ($c_2 \neq 0$) and local doublon-hole
 pairs ($c_1\neq 0$, $\phi=0$) . Fig.~\ref{fig:mzphi} (c) and (d) shows the phase diagram in the $U - V$
plane for the square (c) and the cubic (d) lattice respectively. At
low $U$, there is a first order transition between the AF state and the
SC state, while
at larger  $U$, there is a weakly first order
transition at $V_{c1}$ where the AF order vanishes and a continuous
transition at $V_{c2}$ to the SC phase.
 The later transition should be in the $XY$ universality class~\cite{paired_SF1}.

{\bf Gauge Transformations and the phases:} 
The projection of the slave-boson states to the physical Hilbert space through local constrains
leads to the following gauge invariance of the ionic Hubbard model:
a $U(1)_+$ gauge transform, under which $f_{iA\sigma}\rightarrow
f_{iA\sigma} e^{i\theta_+(i)}, d_{iA}\rightarrow
d_{iA} e^{i\theta_+(i)}, f_{iB\sigma}\rightarrow
f_{iB\sigma} e^{i\theta_+(i)}, h_{iB}\rightarrow
h_{iB} e^{i\theta_+(i)}$ and a $U(1)_-$ invariance, under which $f_{iA\sigma}\rightarrow
f_{iA\sigma} e^{i\theta_-(i)}, d_{iA}\rightarrow
d_{iA} e^{i\theta_-(i)}, f_{iB\sigma}\rightarrow
f_{iB\sigma} e^{-i\theta_-(i)}, h_{iB}\rightarrow
h_{iB} e^{-i\theta_-(i)}$. We note that for charged fermions, under
electromagnetic gauge transformations, the doublons and holes
transform according to $U(1)_-$ gauge due to
their opposite charges.

The doublon-hole pairing and the spinon singlet
amplitude have charge $2$ with respect to the $U(1)_+$
transformations and their finite expectation leads to gapping out the
$U(1)_+$ gauge fields throughout the phase diagram. The current
corresponding to the $U(1)_-$ fluctuations is 
\bqa
\vec{j}^p_-&=& -it \sum_{\langle ij\rangle}
 \vec{r}_{ij}\left[c_2 d^\dagger_{iA}h^\dagger_{jB}+c_1\sigma
 f^\dagger_{iA\sigma}f^\dagger_{jB\overline{\sigma}}-h.c\right]
\eqa
  where $\vec{r}_{ij}=\vec{r}_i-\vec{r}_j$. In the PI phase, the
  current response is given by $\chi^{j_\alpha
  j_\beta}+D^{\alpha\beta}$, with
$D^{\alpha\beta}=-t\sum_{\langle
  ij\rangle}\vec{r}_{ij}^\alpha \vec{r}_{ij}^\beta [ c_2d^\dagger_{iA}h^\dagger_{jB}+c_1\sigma
 f^\dagger_{iA\sigma}f^\dagger_{jB\overline{\sigma}}+h.c.]$. $D^{\alpha\alpha}$
  goes to zero as $q\rightarrow 0,
\omega \rightarrow 0$, leading to a paramagnetic insulator with
neutral vortices consisting of doublons and holes flowing in the same
direction.
The PI phase has deconfined $U(1)_-$ gauge configurations with
associated emergent gapless ``photon'' excitations. Beyond mean-field
and RPA, the deconfined phase can: (i) confine due to instanton
processes in 2D (ii) lead to a $Z_2$ gauge theory due to
intra-sublattice hoppings or (iii) break further lattice
symmetries. In principle, additional terms in the
Hamiltonian can force a single smooth transition from the AF to
the SC phase, where
coupling to the critical matter fields can lead to a stable deconfined
phase for the gauge fields. We note that our dimer-dipole model closely resembles a
model studied earlier by Moessner {\it et. al}~\cite{Moessner}.

In the Bose condensed phase of the doublons/holes, the mode that
condenses is a linear combination of $d^\dagger_k$ and $h_{-k}$ and
has a $U(1)_-$ charge of $1$. The condensation
of this charged mode leads to a superconducting response~\cite{smref1}
with a superfluid
stiffness 
\beq
\rho_s= (2z-4)tc_2\phi^2
\eeq
The superfluid stiffness, plotted as a function of $V$ in
Fig~\ref{fig:spectral}(a), scales with $t$ and shows a non-monotonic dependence on
$V$. As $V$ increases, the condensate fraction $\phi^2$ increases,
while the singlet amplitude $c_2$ decreases, since increase in doublon
density is compensated by decrease in spinon density. The stiffness,
which is a product of these two, thus shows non-monotonic
behaviour. For the superconducting phase, $T_c \sim t$ and will follow the
dome-shape of the stiffness as a function of $V$, reminiscent of the dependence of superconducting $T_c$ of cuprate
superconductors with doping. Further, the destruction of the
superconducting phase due to vortex proliferation at finite
temperatures would lead to a phase with doublon-hole pairing showing
pseudogap behaviour.

The SC phase is characterized by the presence of a linear superposition of
Cooper pairing and $\eta$ pairing~\cite{etapair}, which creates doublons on $A$ and
holes on $B$ sublattice. The vortices in this phase consist of doublons and holes moving in
opposite directions around the vortex core, resulting in charge
currents. In this picture, the charge $1$ of the
vortices has a simple interpretation in terms of charge $2$ objects
(pair of original fermions) flowing through one sublattice, rather
than more exotic topological states~\cite{Subir_vortex}. In 3D, the
presence of both the pair condensate and single particle condensate
will lead to non-trivial drag effects and topological excitations~\cite{sftopo_1,sftopo_2,sftopo_3}.
 
{\bf Single Particle Spectral Function:} The spectral function ${\cal A}(k,\omega)$, which is the probability
density of finding a fermion with a given momentum $k$ and energy
$\omega$ contains detailed information about the single particle
excitations in a system, and is a key measurable quantity both in
material and cold- atom systems. Although the theory is formulated in
terms of spinons and doublons/holes, the measurable quantity is the
spectral function of the original $c$ fermions which is a convolution
of the spectral function of the spinons and the bosons. 

In the AF insulator phase the spinons are gapped on the scale of the
superexchange interaction $\sim t^2/(U+V)$. The kinetic energy, on the
other hand, leads to an extended s-wave
pairing of the spinons ${\cal O}(t)$, where the gap function  has a line node along
the magnetic Brillouin zone. The spectral gap for spinons is shown in
Fig.~\ref{fig:spectral} (b). It goes down with $V$ as AF
order weakens and would have gone to zero if there was a continuous
transition. Instead, a first order transition intervenes, and on the
other side, the spectral gap increases rapidly, driven by the chemical
potential, as in the BEC limit of a BCS-BEC crossover~\cite{BCSBEC}. For the bosons,
the 
quasiparticle spectrum has a minimum at the zone center. The spectral
gap steadily decreases with $V$ till it reaches zero at $V_{c2}$ and
remains zero in the condensed phase. The gap for the $c$ fermions is a
sum of the two gaps and remains finite, as shown in the inset of
Fig.~\ref{fig:spectral}(b). The gap is non-monotonic, dominated by the
bosonic gap at small $V$ and by the spinon gap near $V\sim U$, with a
minimum around $V_{c1}$.

In Fig.~\ref{fig:spectral} (c) and (d), we plot ${\cal A}(k,\omega)$ for the square lattice system, at
$k=[\pi/2,\pi/2]$, which corresponds to the minimum gap point in
this case. In the AF and PI phase, the spectral function
is completely incoherent, while the condensate leads to a coherent
piece of the spectral function, with a residue proportional to the
condensate fraction. The appearance of coherence peaks in the single
particle spectral function can then be used to track the
superconducting transition in this system experimentally. 

{\bf Conclusion:} We have studied the ionic Hubbard model on the
square and cubic lattice, in the limit
of large $U$ and large $V$, when $V \sim U$. Using a low energy
dimer-dipole model and slave
boson mean-field theory, we find that the AF order weakens with
increasing $V$ and vanishes at a critical $V_{c1}$. At larger $V\sim
U$, the system becomes a superconductor, driven by condensation of
charged doublons and holes. This state is characterized by a
coherent but gapped spectral function and a superfluid stiffness,
which is non-monotonic as a function of $V$. This state, which
has a dome shaped $T_c \sim t$,  will also show pseudogap
behaviour as temperature is raised above $T_c$. At large $U$, there is
a paramagnetic insulating phase between the AF insulator and the
superconductor, which, within the mean-field theory, is a gauge
deconfined phase with its associated gapless excitations. This phase
can be understood as a paired superfluid phase of doublons and holes.
In ultracold atomic systems, the superfluid phase is easily detectable
either through single particle spectral function measurements, or
through quantum microscope, which can directly measure the delocalization of
doublons and holes in real space.

\begin{acknowledgements}
The authors thank S. Silotri, K. Damle, H. R. Krishnamurthy,
M. Randeria and S. Sachdev for useful discussions.
\end{acknowledgements}

\bibliographystyle{unsrt} \bibliography{ihm_ref.bib}

\begin{thebibliography}{10}
\bibitem{footnote1}
Placing holes on ${B}$ sublattice instead of ${A}$ sublattice gives an energy
  gain of ${V\sim U}$. this constraint will hold both at half-filling and low
  hole doping, until the system cannot sustain all the holes on the ${B}$
  sublattice in the highly hole doped regime, which we do not consider here.
\end{thebibliography}

\pagebreak


\def\p{\partial}
\def\d{\dagger}
\def\l{\left}
\def\r{\right}
\def\mc{\mathcal}
\def\sl{\sum\limits}
\def\ul{\underline}
\def\ua{\uparrow}
\def\da{\downarrow}
\def\ra{\rightarrow}
\def\Ra{\Rightarrow}
%

\begin{widetext}
\begin{center}
\textbf{\large  Supplementary Material for:  Superconductivity from Doublon
  Condensation in Ionic 
Hubbard Model}
\end{center}
\end{widetext}

%

\maketitle

\section{Canonical Transformation and Low Energy Effective Hamiltonian}
 In a system, whose Hilbert space is fragmented into sectors of energy
 width $\sim \omega_l$ separated by a large energy scale $\omega_h$, a canonical
 transformation can be used to obtain the effective low energy
 Hamiltonian in each sector. The effective Hamiltonian is given by
\begin{eqnarray}
 \tilde H &=& e^{iS}He^{-iS}= H + [iS,H] + \frac{[iS,[iS,H]]}{2!} + .... 
\end{eqnarray}
where $iS$ has a strong coupling perturbation series in $\omega_l/\omega_h$
and is chosen 
in a way that the transformed Hamiltonian $\tilde{H}$ does not have 
any term connecting states belonging to different sectors order by order in $\omega_l/\omega_h$.
The ionic Hubbard Hamiltonian
is given by $H=H_0+H_T$,
\bqa
\no H_0&=&U\sum_i n_{i\up}n_{i\dn} +\frac{V}{2}\sum_i (-1)^i n_i\\
H_T&=&-t\sum_{\langle ij\rangle\sigma}c^\dagger_{i\sigma}c_{j\sigma}=
\sum_nT^n_{-}+T^n_{+}
\eqa
Here $H_0$ is the local part of the Hamiltonian, which includes
both the Hubbard repulsion and the ionic potential. On the $A$
sublattice, single fermion states have energy $-V/2$, doublons have
energy $U-V$ and holes have $0$ energy. On the $B$
sublattice, single fermion states have energy $V/2$, doublons have
energy $U+V$ and holes have $0$ energy. $T^n_{+(-)}$ hops a Fermion
 from the $A(B)$ to the $B(A)$ sublattice and
increases the double occupancy by $n$, with $n=0,\pm 1$. The action of
$T^n_{s}$ on a configuration causes an energy change $\Delta
E^n_s=nU+Vs$, which can be written as $[H_0,T^n_s]=\Delta E^n_s T^n_s$.

In the resonant regime, $U-V$ is a low energy, while $U+V$ and $V$ are
large energies. So the strong coupling expansion is obtained in powers
of $t/(U+V)$ and $t/V$. In this case, it is clear that ${H_T}^l =
\big(T_-^1 + T_+^{-1}\big)$ are low energy hopping terms causing
energy change $\sim U-V$, while the terms ${H_T}^h = {H_T} - {H_T}^l$
should be eliminated by canonical transformation. This is achieved by 
 \begin{equation}
iS^{(1)} = \frac{1}{U+V}\l(T_+^1-T_-^{-1}\r) 
+ \frac{1}{V}\l(T_+^0-T_-^0\r)  
 \end{equation}
Out of the second order terms generated, it is clear that products of
the form $T^n_sT^{-n}_{\overline{s}}$ brings the system back to the
sector it started from, and would survive in second order, while the
rest should be eliminated. This is achieved by 
\begin{widetext}
\bqa
\no
iS^{(2)}&=&\frac{1}{2(U+2V)}\left(\frac{1}{U+V}-\frac{1}{V}\right)\left([T^1_+,T^0_+]+[T^{-1}_-,T^0_-]\right)+\frac{1}{2U}\left(\frac{1}{U+V}+\frac{1}{V}\right)\left([T^1_+,T^0_-]+[T^{-1}_-,T^0_+]\right)\\
& &
+\frac{1}{2U}\frac{1}{U+V}\left([T^1_+,T^1_-]+[T^{-1}_-,T^{-1}_+]\right)+\frac{1}{2V}\frac{1}{U+V}\left([T^{-1}_-,T^1_-]+[T^{1}_+,T^{-1}_+]\right)\\
\no& &
-\frac{1}{UV}\left([T^1_-,T^0_+]+[T^{-1}_+,T^0_-]\right)+\frac{1}{V(U-2V)}\left([T^1_-,T^0_-]+[T^{-1}_+,T^0_+]\right)
\eqa
The effective low energy Hamiltonian upto second order ($t^2/U$) in the 
perturbation expansion takes the following form :
 \bqa
 \no \tilde H &=& H_0 +H_T+[iS^{(1)},H_0+H_T]+[iS^{(2)},H_0]+\frac{1}{2}[iS^{(1)},[iS^{(1)},H_0]]\\
&=& H_0 +\l(T_-^1 + T_+^{-1}\r) + \frac{1}{U+V} \l[T_+^1,T_-^{-1}\r] 
 +\frac{1}{V} \l[T^0_+,T^0_-\r]
 \eqa
\end{widetext}
\section{ Effective Hamiltonian Using Slave Boson Operators }
 In the slave boson formalism, the fermion operator $c^\dagger_{i\sigma}$ is 
 written in
 terms of spinful fermions (spinons) $f^\dagger_{i\sigma}$ and charged
 bosons, doublons $d^\dagger_{i}$ with charge $+1$ and holons
 $h^\dagger_i$ with charge $-1$. This is given by
 $c^\dagger_{i\sigma}=f^\dagger_{i\sigma}h_i+\sigma f_{i\overline{\sigma}}d^\dagger_i$
 along with the constraint equation
 $f^\dagger_{i\sigma}f_{i\sigma}+d^\dagger_id_i+h^\dagger_ih_i=1$. In
 the resonant regime, configurations with doublons on $B$ sublattice
 and holons on $A$ sublattice are projected out of the low energy
 subspace and hence we can write
 $c^\dagger_{iA\sigma}=\sigma f_{iA\overline{\sigma}}d^\dagger_{iA}$ and
 $c^\dagger_{iB\sigma}=f^\dagger_{iB\sigma}h_{iB}$. The constraint
 equations are then given by
 $f^\dagger_{iA\sigma}f_{iA\sigma}+d^\dagger_{iA}d_{iA}=1$ and 
 $f^\dagger_{iB\sigma}f_{iB\sigma}+h^\dagger_{iB}h_{iB}=1$. Using this, we can write 
 different low energy hopping terms in the Hamiltonian in the
 following way :
 \\
\begin{figure}[t]
\includegraphics[width=\linewidth]{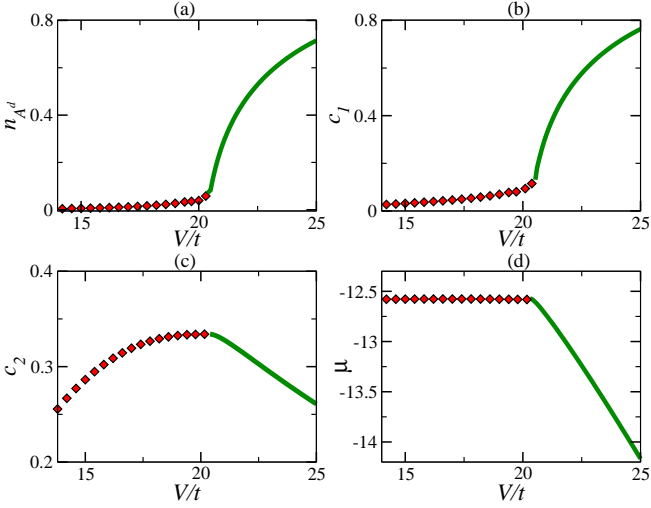}
\caption{Different Mean field parameters as a function of the ionic potential $V$ 
for a cubic lattice 
with $U=25t$ (a) doublon number density ($n_A^d$) (b) doublon-hole pairing ($c_1$)
(c) spin singlet pairing ($c_2$) (d) chemical potential ($\tilde\mu$).}
\label{fig:othermean}
\end{figure}

 \ul{$\mc{O}(t)$ hopping terms} :
 \begin{eqnarray}
 T^1_- +T^{-1}_+ &=&-t\sum\limits_{\langle ij\rangle\sigma} 
 \sigma f_{jB\bar{\sigma}} f_{iA\sigma} d_{iA}^\dagger h_{jB}^\dagger  + h.c. 
 \end{eqnarray}
 It is clearly seen that the low energy hopping process is equivalent
 to a spin singlet on a bond fluctuating to a doublon-holon pair (a
 charge dipole) and vice-versa.
\\

 \ul{ $\mc{O}(t^2/U)$ terms involving sites on a single bond} :
\begin{eqnarray}
 \frac{1}{U+V}\l[T^1_+,T^{-1}_-\r] &=& \frac{2t^2}{U+V}\sl_{\langle ij\rangle}
 \vec{S}_i\cdot \vec{S}_j-\frac{1}{4} n^f_in^f_j
\end{eqnarray}
This is the usual superexchange term with a reduced superexchange
scale $2t^2/(U+V) $. There is a factor of $2$ reduction since the
intermediate virtual doublon can only be created on the $B$ sublattice
(doublons on $A$ sublattice are low energy configurations). The other
term is given by 
\begin{eqnarray}
 \frac{1}{V}\l[T^0_+,T^0_-\r] 
 &=& -\frac{ t^2}{V}\sl_{\langle ij\rangle\sigma}\l[n_{iA}^\sigma n_{jB}^h 
 + n_{iA}^d n_{jB}^{\bar\sigma} \r]
\end{eqnarray}
 The attractive interaction between doublons/holes and spinons can
 alternatively be thought of as an effective repulsion between
 neighbouring doublons and holes by using the constraint equations to
 eliminate the spinons in these terms.
 \\ \ul{Intra-sublattice $\mc{O}(t^2/U)$ hopping terms} :
These terms are given by
 \begin{eqnarray}
 & &\frac{t^2}{V}\sl_{\langle ijl\rangle\sigma}
 \Big[\l(f^\dagger_{lA\sigma}f_{iA\sigma}n_{jB}^{\sigma}
 + f^\dagger_{lA\sigma}f_{iA\bar\sigma}f^\dagger_{jB\bar\sigma}
 f_{jB\sigma}\r)d^\dagger_{iA}d_{lA} \no\\
 && + \l(f^\dagger_{lB\sigma}f_{iB\sigma}n_{jA}^{\sigma}
 + f^\dagger_{lB\sigma}f_{iB\bar\sigma}f^\dagger_{jA\bar\sigma}f_{jA\sigma}\r)
 h^\dagger_{iB}h_{lB}\Big]
 \end{eqnarray}
 These terms describe the hopping doublon (holon) from a site to 
 its next nearest neighbour site on same sublattice with an associated
 backflow of spinons. Note that the
 absence of doublons on $B$ sublattice and holes on $A$ sublattice
 precludes the possibility of a $t^2/(U+V)$ term of this kind. In
 addition to these terms, we use Lagrange multipliers $\mu_i^A$ and
 $\mu_i^B$ to implement the constraint on te two sublattices. In all
 the effective Hamiltonian can be written as
 \begin{widetext}
 \begin{eqnarray}
 \tilde H& =& \sum_i
 \l[(U-V-2\mu-\mu_i^A)n^d_{iA}-\mu_i^B n^h_{iB}\r]-\sum_i
 \l[\l(V/2+\mu+\mu_i^A\r)n^f_{iA} + \l(\mu+\mu_i^B-V/2\r)n^f_{iB}\r] \no\\
 & &-t \sum_{\langle ij\rangle \sigma} \sigma
 \l(f_{jB\overline{\sigma}}f_{iA\sigma}d^\dagger_{iA}h^\dagger_{jB}+h.c.\r)
 + \frac{t^2}{U+V}\sum_{\langle ij\rangle\sigma} 
 \l( -n_{iA}^\sigma n^{\bar\sigma}_{jB}+ f_{iA\bar\sigma}^\dagger 
 f_{iA\sigma}f_{jB\sigma}^\dagger f_{jB\bar\sigma}\r)
 - \frac{t^2}{V}\sl_{\langle ij\rangle\sigma}\l(n_{iA}^\sigma n_{jB}^h 
 + n_{iA}^d n_{jB}^\sigma\r)  \no\\
 &+& \frac{t^2}{V} \sl_{\langle ijl\rangle\sigma}\Big[\l(f^\dagger_{lA\sigma}
 f_{iA\sigma}n_{jB}^{\sigma}
 + f^\dagger_{lA\sigma}f_{iA\bar\sigma}f^\dagger_{jB\bar\sigma}f_{jB\sigma}\r)
 d^\dagger_{iA}d_{lA} +\l(f^\dagger_{lB\sigma}f_{iB\sigma}n_{jA}^{\sigma}
 + f^\dagger_{lB\sigma}f_{iB\bar\sigma}f^\dagger_{jA\bar\sigma}f_{jA\sigma}\r)
 h^\dagger_{iB}h_{lB}\Big]
\end{eqnarray}
\end{widetext}

\section{ Mean Field Theory} 

We use a mean field theory where we decouple the bosons 
and fermions and the constraints are maintained on average.
We take the following mean field parameters : the staggered magnetization 
$ m_s = \frac{1}{2}\langle S_{iA}^z-S_{iB}^z\rangle$ where $S^z = \frac{1}{2}
\langle f_{\up}^\dagger f_{\up} - f_{\dn}^\dagger f_{\dn} \rangle$, 
the singlet pairing amplitude
$c_2=\langle \sigma f_{iA\sigma}f_{jB\bar\sigma}\rangle$, 
the doublon-holon pairing amplitude $c_1 = \langle
d_{iA}h_{jB}\rangle$. In addition we consider intra-sublattice hopping
amplitude for both spinons and the bosons,
$c_3=\langle f_{iA\sigma}^\dagger f_{lA\sigma}\rangle$ 
and $c_4=\langle d_{iA}^\dagger d_{lA}\rangle$. 
We note that individual spinon and boson densities are
determined by the chemical potential and the Lagrange multipliers
which implement the constraints. We have checked that out of different
spatial symmetries for the pairing, the system always chooses extended
s-wave pairing on energetic grounds.

At half-filling, charge neutrality forces equality between the doublon
density $n^d$ and the holon density $n^h$. The constraint equations
then imply that $n^f_A=n^f_B$, i.e. the spinon density on the two
sublattices are equal. This gives us :
 $(U-V)-2\mu-\mu^A=-\mu^B$ and
 $-\mu-\mu^A-\frac{V}{2}=-\mu-\mu^B+\frac{V}{2}$, 
i.e.  $\mu=\frac{U}{2}$, 
 and $\mu^B-\frac{V}{2}=\mu^A+\frac{V}{2}=\tilde\mu$. This gives us a
 bosonic Hamiltonian
\begin{equation}
H^B=\sum\limits_{k} \left( \begin{array}{cc}
d_{kA}^\dagger & h_{-kB}    \end{array} \right) 
\left( \begin{array}{cc}
a^B_{k}& -t c_2\gamma_k \\
-t c_2 \gamma_k & a^B_{k}  \end{array} \right)
\left( \begin{array}{c}
d_{kA} \\
h_{-kB}^\dagger  \end{array} \right)
\end{equation}
where, $a^B_{k}=-\tilde\mu - \frac{V}{2} - zJ_3 n_A^f
+ \frac{1}{2}J_3(\gamma_k^2-z).(c_3n_A^f-c_2^2)$, where $J_1=\frac{t^2}{U+V}$ 
and $J_3=\frac{t^2}{V}$ and $\gamma_k=2\sl_{\alpha}\cos k_\alpha $. 
The Hamiltonian for fermions can be written as
\begin{equation}
H^F=\sum\limits_{k,\sigma}   \left( \begin{array}{cc}
f_{kA\sigma}^\dagger & f_{-kB\bar\sigma}  \end{array} \right) 
\left( \begin{array}{cc}
a^F_{k\sigma}& c^F_{k\sigma} \\
c^F_{F\sigma}& -a_{kF\sigma}\end{array} \right)
\left( \begin{array}{c}
f_{kA\sigma} \\
f_{-kB\bar\sigma}^\dagger \end{array} \right)
\end{equation}
 where, $a^F_{k\sigma}= -\tilde\mu -\frac{U}{2} - \frac{1}{2}zJ_1 n_A^f
 -zJ_1\sigma m_s - zJ_3n^d +\frac{1}{2}J_3(\gamma_k^2-z)n_B^f c_4 
 +\frac{1}{2}z(z-1)J_3c_3c_4$
 and $c^F_{k\sigma}=-t\sigma\gamma_kc_1 -\frac{1}{2}\sigma\gamma_k
 J_1 c_2 -J_3\sigma \gamma_k(z-1)c_2 c_4$.

The bosonic Hamiltonian is diagonalized by a
Bogoliubov transformation to yield the spectrum $E^B_{k} = \pm
\sqrt{(a^B_{k})^2 - t^2c_2^2\gamma_k^2}$, while the fermionic spectrum is
given by $E^F_{k\sigma}= \pm
\sqrt{(a^F_{k\sigma})^2 +( c^F_{k\sigma})^2}$. If the bosonic spectrum
reaches $0$ at any point on the Brillouin zone for a set of parameters, 
a condensate of the
corresponding quasiparticles would form at that point. We have found
that for s-wave symmetry of the doublon-hole pairing, this always
occurs at the zone center. In that case, half-filling dictates that
$\langle d^\dagger(0,0)\rangle=\langle h^\dagger(0,0)\rangle=\phi$ in 2D and
$\langle d^\dagger(0,0,0)\rangle=\langle h^\dagger(0,0,0)\rangle=\phi$ in 3D,
which the off-diagonal doublon-holon pairing terms fixing the relative
phase of the condensate. In the condensed phase of the doublons and holons, the
mean-field equations are
{\begin{widetext}
 \begin{gather}
 n_A^d = \phi^2 +\frac{1}{N}\sl_{k'}\l[(u_k^B)^2 n_B(E_k^B) 
 + (v_k^B)^2 (1+n_B(E_k^B))\r] \no\\
 m_s = \frac{1}{2N}\sl_{k,\sigma}\sigma \l[(u_{k\sigma}^F)^2 n_F(E_{k\sigma}^F) 
 + (v_{k\sigma}^F)^2 \l(1 - n_F(E_{k\sigma}^F)\r)\r] \no\\
 c_1 = \phi^2 -\frac{1}{zN}\sl_{k'}\gamma_ku_k^Bv_k^B
 \l[1 + 2n_B(E_k^B) \r] \no\\
 c_2 = \frac{1}{zN}\sl_{k,\sigma} \bar\sigma\gamma_k 
 u_{k\sigma}^Fv_{k\sigma}^F
 \l[1 - 2n_F(E_{k\sigma}^F)\r] \no\\
 c_3 =  \frac{1}{z(z-1)N}\sl_{k,\sigma} (\gamma_k^2-z)\l[(u_{k\sigma}^F)^2 
 n_F(E_{k\sigma}^F) + (v_{k\sigma}^F)^2 \l(1 - n_F(E_{k\sigma}^F)\r)\r] \no\\
 c_4 = \phi^2 +\frac{1}{z(z-1)N}\sl_{k'}(\gamma_k^2-z)
 \l[(u_k^B)^2 n_B(E_k^B) + (v_k^B)^2 (1+n_B(E_k^B))\r] \no\\
 n_A^d + \frac{1}{N}\sl_{k,\sigma}\l[(u_{k\sigma}^F)^2 n_F(E_{k\sigma}^F) 
 + (v_{k\sigma}^F)^2 \l(1 - n_F(E_{k\sigma}^F)\r)\r] = 1
\end{gather}
 \end{widetext}}
where $k'$ are the all other $k$ points except where condensation occurs
and $n_B$ and $n_F$ are Bose function and Fermi function respectively. 
Bosonic coherence factors $u^B_k$ and $v^B_k$ are given by
$(u_k^B)^2=\frac{1}{2}\l( 1 + \frac{a_k^B}{E_k^B} \r)$ and 
$(v_k^B)^2=\frac{1}{2}\l( \frac{a_k^B}{E_k^B} -1 \r)$ respectively and the 
fermionic coherence factors $u^F_{k\sigma} $ and $v^F_{k\sigma} $ are given by 
$(u_{k\sigma}^F)^2=\frac{1}{2}\l( 1 + \frac{a_{k\sigma}^F}{E_{k\sigma}^F} \r)$ and 
$(v_{k\sigma}^F)^2=\frac{1}{2}\l( 1 - \frac{a_{k\sigma}^F}{E_{k\sigma}^F} \r)$ 
respectively. The equations for the uncondensed phase can be obtained
by setting $\phi=0$. In the condensed phase, we
use the additional equation $E^B_{k}(0,0)=0$ in 2D and $E^B_{k}(0,0,0)=0$ in 3D 
ensuring gaplessness of Goldstone modes.

The main features of the mean-field phase diagram together with the evolution
of the order parameters $m_s$ and $\phi$ as a function of $V$ is
discussed in the main text. The evolution of the other mean fields are
shown in Fig.~\ref{fig:othermean} for a cubic lattice with
$U=25t$. The dotted points are in the phase where doublons are not
condensed, whereas the condensed phase is shown with solid line. Fig.
~\ref{fig:othermean}(a) and (b) show the doublon density and the
doublon-holon pairing amplitude $c_1$. Both these quantities show a
rapid rise in the condensed phase, with the transition point providing
a point of inflection. Fig. ~\ref{fig:othermean}(c) and (d) show the
spinon pairing amplitude and the chemical potential. The spinon
pairing increases as the system moves away from AF order and then decreases
in the condensed phase as rapid increase of doublons force a smaller
density of spinons in the system. The chemical potential also shows
rapid changes with $V$ in the condensed phase, reflecting the changes
in charged degrees of freedom.

\section{Superfluid Stiffness }
Within the slave boson mean field theory, the kinetic energy term
coupled to a $U(1)_-$ gauge field is given 
by 
{\small
\begin{eqnarray}
 H_T&&\ra H_T(\vec{A}) = 
 -tc_2 \sl_{\langle ij\rangle} 
 \l( d^\d_{iA} h^\d_{jB}e^{-i\vec{A}.(\vec{r_i}-\vec{r_j})} + h.c. \r) \no\\
 &&-tc_1 \sl_{\langle ij\rangle\sigma} \sigma\l( f^\d_{iA\sigma} 
 f^\d_{jB\bar\sigma}e^{-i\vec{A}.(\vec{r_i}-\vec{r_j})} + h.c. \r)
\end{eqnarray}
}
where, $\vec{A}$ is the corresponding vector potential. The
paramagnetic current on a bond between $\vec{r}_i$ and 
$\vec{r}_j$ is 
\begin{eqnarray}
\vec{j^p_-}(q) &=& tc_2\sl_{k} \l( \rho_{k+\frac{q}{2}}d^\d_{kA}h^\d_{-k-q,B} + h.c.\r) \no\\
 +&& tc_1\sl_{k,\sigma}\sigma \l( \rho_{k+\frac{q}{2}}
 f^\d_{kA\sigma}f^\d_{-k-q,B\bar\sigma} +h.c.\r)
\end{eqnarray}
with $\rho_k=2\sl_{\alpha}\sin{k_\alpha}$. 
The response of the system is given by 
\begin{equation}
  \langle (\vec{j_-^p}(q))_\alpha 
 (\vec{j_-^p}(-q))_\beta \rangle+D^{\alpha\beta}(q)
 \end{equation}
where the diamagnetic response is given by 
\bqa
D^{\alpha\beta}(q\rightarrow 0)&=& t\sum_k \frac{\partial^2
  \gamma_k}{\partial k_\alpha\partial k_\beta} \Big[ c_2 \l(\langle
d^\dagger_{kA}h^\dagger_{kB} \rangle + h.c.\r)\no\\
& & + c_1\sl_{\sigma}\sigma\l( \langle
f^\dagger_{kA\sigma}f^\dagger_{kB\bar{\sigma}}\rangle + h.c.\r) \Big]
\eqa

In the non-condensed phase, the paramagnetic response is  
\begin{eqnarray}
 \chi^{\alpha\alpha}(0)&=& t^2 \sl_{k} \l(\frac{\p{\gamma_k}}{\p k}\r)^2_\alpha \l[
 c_2^2 \frac{(a^B_k)^2}{(E^B_k)^3} +c_1^2\frac{(a^F_{k\sigma})^2}{(E^F_k)^3} \r] 
 \end{eqnarray}
while the diamagnetic response is given by 
\begin{eqnarray}
 D^{\alpha\alpha}&=& -2t \sl_{k} \frac{\p^2{\gamma_k}}{\p k_\alpha^2} \l[
 c_2 u^B_kv^B_k +c_1\sl_{\sigma}\sigma u^F_{k\sigma}v^F_{k\sigma} \r] 
 \end{eqnarray}
Using integration by parts, it can be easily shown that the paramagnetic and diamagnetic responses cancel each 
other exactly in the $q\rightarrow 0$ limit and hence the superfluid
stiffness is $0$ in this phase. Note that the spinon and boson
contributions cancel individually, and the system is an insulator.

In the condensed phase, the above calculation goes through, except for
the fact that the  condensation of the bosons add a new term to the
current of the form 
\begin{equation}
\delta \vec{j_{B-}^p}(q)
 = -tc_2 \rho_{\frac{q}{2}} [ \phi^*(d^\d_{-qA} + h_{qB})
 -\phi(h^\d_{-qB} + d_{qA}) ] 
 \end{equation}
Working out the paramagnetic current-current correlator from this
additional term we get
\begin{eqnarray}
 \delta \chi^{\alpha \alpha}&=& 4tc_2|\phi|^2   \label{condpar} 
\end{eqnarray}
Similarly, the diamagnetic response gets an additional term 
\begin{eqnarray}
 \delta D^{\alpha\alpha}  &=& -2ztc_2|\phi|^2 \label{conddia}
\end{eqnarray}
Combining eqn.(\ref{condpar}) and (\ref{conddia}), we see that the contribution from 
paramagnetic and diamagnetic responses in the condensed phase do not cancel
each other and  we obtain the finite superfluid stiffness, 
$\rho_s=(2z-4)tc_2|\phi|^2$.

\section{Single Particle Spectral Functions }
The single particle spectral function $\mc{A}(k,w)$, which is proportional to the 
imaginary part of the Green’s function, gives the probability density that a 
particle with a certain momentum $k$ has a specific energy $\omega$. The key 
point to note is that the original $c$ fermions constitute gauge invariant 
operators and hence their single particle Green’s functions will be measured 
by different experiments like ARPES, STS etc. Since the $c$ fermions are written as 
product of $f$ fermions and d/h bosons, the single particle Green’s function 
calculation for the $c$ fermions is akin to a bubble calculation of polarization 
function, with one fermion and one boson line forming the bubble. 
For example, the single particle fermion Green's function for a up-spin fermion
on $A$ sub-lattice is defined as : 
\begin{eqnarray}
&&\mc{G}_{AA}^{\ua}(i,j,\tau,\tau')
=-T_{\tau\tau'}\l<c_{iA\ua}(\tau)c^\d_{jA\ua}(\tau')\r> \no\\
&=& -T_{\tau\tau'}\l[\l<f^\d_{iA\da}(\tau)f_{jA\da}(\tau')\r>
\l<d_{iA}(\tau)d^\d_{jA}(\tau')\r>\r] \no\\
&=& D_{11}(\tau,\tau')G^{\dn}_{11}(\tau',\tau)
\end{eqnarray}
where, the angular bracket denotes the expectation value,
$T$ is the time ordering operator, and
the boson and spinon Green’s functions are given by
\begin{widetext}
\begin{equation}
D(k,i\omega_n)=\left(\begin{array}{cc}
  \frac{(u_k^B)^2}{i\omega_n-E_k^B}-\frac{(v_k^B)^2}{i\omega_n+E_k^B}
& 
-u_k^B v_k^B\l(\frac{1}{i\omega_n-E_k^B}-\frac{1}{i\omega_n+E_k^B}\r) \\ \\ 
-u_k^B v_k^B\l(\frac{1}{i\omega_n-E_k^B}-\frac{1}{i\omega_n+E_k^B}\r) 
& 
\frac{(v_k^B)^2}{i\omega_n-E_k^B}-\frac{(u_k^B)^2}{i\omega_n+E_k^B }
  \end{array}
 \right)	
 \end{equation}		
\begin{equation}
G^\sigma(k,i\omega_n)=\left(
  \begin{array}{cc}
  \frac{(u_{k\sigma}^F)^2}{i\omega_n-E_{k\sigma}^F}+\frac{(v_{k\ua\sigma}^F)^2}
  {i\omega_n+E_{k\sigma}^F}
& 
-u_{k\sigma}^F v_{k\sigma}^F\l(\frac{1}{i\omega_n-E_{k\sigma}^F}-\frac{1}{i\omega_n+E_{k\sigma}^F}\r) \\ \\ 
-u_{k\sigma}^F v_{k\sigma}^F\l(\frac{1}{i\omega_n-E_{k\sigma}^F}-\frac{1}{i\omega_n+E_{k\sigma}^F}\r)
& 
\frac{(v_{k\sigma}^F)^2}{i\omega_n-E_{k\sigma}^F}+\frac{(u_{k\sigma}^F)^2}{i\omega_n+E_{k\sigma}^F}
  \end{array}
\right)				\no\\
\end{equation}
\end{widetext}
Fourier transforming to the momentum and Matsubara frequency space we get 
\begin{equation}
 \mc{G}_{AA}^{\ua}(k,i\omega_n)=\sl_{q}\frac{1}{\beta}
 \sl_{iq_l}G_{11}^\da(q,iq_l)D_{11}(k+q,i\omega_n+iq_l ) 
\end{equation}
where $q_l=(2l+1)\pi T$, with integer $l$, is the fermionic Matsubara frequency. Working
out the Matsubara sum, and taking the analytical continuation
$i\omega_n \rightarrow\omega+i0^+$, we get the spectral function
\begin{eqnarray}
{\mc{A}_{AA}^\ua}(k,\omega) = \sl_{q} && \int\frac{d\omega'}{\pi^2} 
\l[n_F(\omega') + n_B(\omega'+\omega)\r] \no\\
&& .(G^\dn_{11})''(q,\omega')D_{11}''(q+k,\omega+\omega')
\end{eqnarray}
where $''$ denotes imaginary part of the Green's function.
Specializing to $T=0$, we get 
\begin{eqnarray}
 &&{\mc{A}_{AA}^\ua}(k,\omega) \no\\
 &&= \sl_{q}  (u_{k+q}^B)^2 \Big[(v_{q\ua}^F)^2\delta(\omega -E_{q\ua}^F
 -E_{k+q}^B) \Theta(\omega-E_{q\ua}^F) \no\\
 && + (u_{q\ua}^F)^2\delta(\omega +E_{q\ua}^F +E_{k+q}^B) 
 \Theta(|\omega|-E_{q\ua}^F)\Big] 
\end{eqnarray}
In the bose condensed phase of the doublons, the spectral function
picks up an additional contribution from the condensate given by 
\begin{eqnarray}
 {\mc{A}_{AA}^\ua}(k,\omega) 
 &=&-\frac{1}{\pi} \phi^2 (G^\dn_{11})''(k,\omega) \\
 &=& \phi^2\l[ (u_{k\ua}^F)^2 \delta(\omega -E_{k\ua}^F) + (v_{k\ua}^F)^2 
 \delta(\omega +E_{k\ua}^F)\r] \no
\end{eqnarray} 

So, the zero temperature spectral function for fermion on $A$ sub-lattice are given
as following :  
\begin{widetext}
\begin{gather}
 \mc{A}_{AA}(k,\omega) = \sl_{q,\sigma} \Big[(v_{q\sigma}^F)^2 (u_{k+q}^B)^2 
 \delta(\omega -E_{q\sigma}^F-E_{k+q}^B)\Theta(\omega-E_{q\sigma}^F)\no\\ 
 + (u_{q\sigma}^F)^2 (v_{k+q}^F)^2 \delta(\omega +E_{q\sigma}^F +E_{k+q}^B)
 \Theta(|\omega|-E_{q\sigma}^F)\Big]
 + \phi^2\sl_{\sigma}\l[ (u_{k\sigma}^F)^2 \delta(\omega -E_{k\sigma}^F) 
 + (v_{k\sigma}^F)^2 \delta(\omega +E_{k\sigma}^F)\r] \no\\ 
 \label{spec}
 \end{gather}
 \end{widetext}
Similarly, we can calculate $\mc{A}_{BB}(k,\omega)$ and 
$\mc{A}_{AB}(k,\omega)$. The full single particle spectral function is 
given by $\mc{A}(k,\omega)
= \mc{A}_{AA}(k,\omega) + \mc{A}_{BB}(k,\omega)
+ \mc{A}_{AB}(k,\omega)+\mc{A}_{BA}(k,\omega)$. It is clear that in
the non-condensed phase the convolution gives rise to an incoherent
spectral function. The gauge fluctuations, which will provide vertex
corrections, will not change the incoherent nature of the spectral
function. On the other hand, the condensate provides a coherent part
to the spectral function, which is simply proportional to the spinon
spectral function.
Here, we see that the fermion spectral function has peaks at 
$|\omega|=E_{q\sigma}^F+E_{k+q}^B$, which corresponds to the gap of the spectral 
function and the minimum gap is given by 
$\omega_c=E_{k\sigma}^F(\pi/2,\pi/2)+E_k^B(0,0)$. 



\end{document}